# Subsurface Propagation Characteristics of Medium-Wave Electromagnetic Fields Revealed by Measurements in the Nanatsuo-guchi Quarry: Conceptual Framework of the Subground Wave and the Rainfall Model

Eiichi Shoji*

† *Advanced Materials Innovation & Monozukuri Lab, University of Fukui, 3-9-1 Bunkyo, Fukui 910-8507, JAPAN*

*Email: shoji@u-fukui.ac.jp, phone: +81-776-27-8076

**Abstract**

This paper presents field observations of medium-wave (MW; 300 kHz–3 MHz) radio signals propagating in the subsurface rock environment of the Nanatsuo-guchi quarry, an underground Shakudani Ishi excavation site on Mt. Asuwayama in Fukui City, Japan. MW broadcast signals from a nearby local station (JOFG, 927 kHz, 5 kW), received mainly as a surface wave, and from a distant station (JOAB, 693 kHz, 500 kW), received via ionospheric reflection, were successfully received deep inside the quarry, whereas very-high-frequency frequency-modulated (FM) broadcast signals attenuated rapidly and became undetectable near the entrance. This contrasting behavior highlights the strong wavelength dependence of electromagnetic-wave propagation in subsurface environments. Two-axis rotation measurements were performed using loop antennas to analyze the arrival direction and angular dependence of the received signals. In addition to the horizontal magnetic field component ($H_x$), dominant near the ground surface, a vertical magnetic field component ($H_z$) was consistently observed inside the quarry. The relative contribution of $H_z$ increased with depth and was accompanied by systematic variations in the apparent arrival direction. Inclination measurements further revealed a characteristic minimum in reception sensitivity near 40°–50°, suggesting a composite magnetic field structure involving both $H_x$ and $H_z$. These observations cannot be fully explained by conventional surface-wave propagation models based on the Zenneck–Sommerfeld formulation, and instead suggest the formation of a characteristic electromagnetic field structure under subsurface boundary conditions. This study provides experimental evidence for previously unreported MW field behavior in underground spaces and offers new perspectives for subsurface communication and disaster-resilient information systems.

## 1. Introduction

Medium-wave (MW) band (300 kHz–3 MHz) radio broadcasting has served as an important communication infrastructure since the early twentieth century because long-wavelength electromagnetic waves propagate over long distances along conductive ground surfaces (Collin, 1992; Kraus and Marhefka, 2002). This propagation phenomenon, generally described as surface-wave propagation, is theoretically based on the concept of interface waves (Zenneck waves) proposed by Jonathan Zenneck (Sommerfeld, 1909; Zenneck, 1907). These electromagnetic waves propagate along the interface between air and a conductive medium while gradually undergoing attenuation. However, because the original formulation assumes an ideal conductor, its direct application to an actual ground surface has certain limitations (Wait, 1970). This formulation was later extended by Sommerfeld, who derived an exact solution for media with finite conductivity and established a propagation model incorporating more realistic ground conditions (Norton, 1937; Norton, 1941; Sommerfeld, 1949). This formulation provides a systematic mathematical description of electromagnetic waves radiating from vertical antennas and propagating along the ground surface.

Subsequently, Norton introduced a practical approximation formula that incorporated the conductivity, permittivity, and frequency characteristics of the ground, enabling the quantitative determination of surface wave field strength distributions (Norton, 1937; Norton, 1941). These theoretical developments are now incorporated into the ITU-R Recommendation P.368 and constitute the standard framework for analyzing MW propagation under various ground conditions. The theoretical basis of these models is Maxwell’s equations, which indicate that electromagnetic waves are transverse, and the electric field $E$, magnetic field $H$, and propagation vector $\mathbf{k}$ are mutually orthogonal. The electromagnetic energy flow is described by the Poynting vector $\mathbf{S}$ (= $\mathbf{E} \times \mathbf{H}$). In MW broadcasting, it is generally assumed that vertically polarized waves are radiated from vertical antennas, wherein the dominant component of the electric field is vertical ($E_z$) relative to the ground surface, and the magnetic field is oriented horizontally (Ishimaru, 1991; Wait, 1982). In the present study, polarization was defined based on the orientation of the electric field. The observed vertical magnetic field component ($H_z$) is attributed to a redistribution of the magnetic field under specific boundary conditions, with no change in polarization.

Sommerfeld's exact solution is closely related to a representation in which electromagnetic fields are expressed as superpositions of plane waves. In plane-wave expansion, which was later formalized by Weyl, the electromagnetic field radiating from an antenna can be described as a continuous superposition of plane-wave components at various incident angles. From this viewpoint, surface-wave propagation can be interpreted not as a single-mode phenomenon but as a result of the interference of multiple components. Such representations were discussed in classical electromagnetic theory (Stratton, 1941). Consequently, the direction of the Poynting vector **S** is not necessarily unique and may include vertical components depending on the boundary conditions and field distributions. Thus, electromagnetic energy flow is not strictly constrained to be parallel to the ground surface, although the conventional surface-wave theory assumes rapid attenuation of fields penetrating the ground.

However, these theoretical models were developed primarily for open environments near the ground surface. The electromagnetic behavior in closed subsurface spaces, such as tunnels, caves, and quarries, has not been explicitly considered. Wavelengths in the MW band (approximately 300–1000 m) are significantly larger than the cross-sectional dimensions of typical underground structures, which are generally on the order of several meters. This scale mismatch suggests that electromagnetic wave penetration and propagation in confined environments are limited (King et al., 1992). Accordingly, textbooks generally state that MW signals attenuate significantly underground, and their reception is difficult (Wait, 1970). Although theoretical analyses of propagation in conductive media have been conducted (Wait, 1970), systematic field observations of natural subsurface environments remain scarce (Pozar, 2011). In addition, previous studies have demonstrated that MW broadcast waves can be utilized for energy harvesting over wide geographical ranges and for battery-free radio systems, such as hoop-shaped radio (HOOPRA), highlighting their robustness and practical applicability (Shoji, 2016; 2023a; 2024). Related technologies for MW radio wave energy harvesting and loop-based antenna systems have also been developed and protected by patents(Shoji, 2023b; 2025a; 2025b).

To address this gap, the present study investigated MW propagation in the Nanatsuo-guchi quarry, a historical underground excavation site for Shakudani Ishi, on Mt. Asuwayama, Fukui City, Japan (Figure 1A and 1B). The quarry has a long history and consists of bedrock composed mainly of dacitic pumiceous volcanic lapilli tuff (Yoshizawa, 2008). This rock is homogeneous in composition and contains numerous pores (Yoshizawa, 2016). Under humid conditions, the moisture in the pores can increase the effective conductivity and permittivity of the rock. Inside the quarry, the temperature remains at approximately 13°C throughout the year, and the humidity approaches 100% (Yoshizawa, 2008), resulting in a stable, low-noise electromagnetic environment. Such conditions may stabilize the boundary between the cavity and surrounding bedrock. The cross-sectional dimensions of the excavated space are only several meters, which is extremely small compared to the wavelengths in the MW band. Under such scale differences, the cavity boundary conditions may influence the electromagnetic field distributions. Therefore, the Nanatsuo-guchi quarry provides a real environment with geometric and physical conditions that differ from those of the semi-infinite ground assumed in classical theory. Another important feature of quarries is the presence of transmitters in their vicinity. An MW station (JOFG; 927 kHz; 5 kW) is located approximately 3.7 km east–southeast of the quarry entrance. In addition, three frequency modulation (FM) broadcasting stations were located approximately 500 m away on the ridge of Mt. Asuwayama. This configuration enables the direct investigation of subsurface wave propagation and its wavelength dependence. Field measurements revealed that MW signals could be clearly received and demodulated inside the quarry, whereas very-high-frequency (VHF) band FM signals attenuated rapidly near the entrance and became impossible inside the quarry. This contrasting behavior highlights the wavelength dependency of wave propagation. Furthermore, measurements indicated the presence of the $H_z$ component in the subsurface environment, in addition to the horizontal component ($H_x$) that dominates near the ground surface. The relative contribution of $H_z$ increased with increasing depth, suggesting that the electromagnetic field structure differed from the conventional surface-wave model predictions.

The objective of this study was to experimentally investigate the electromagnetic field structure of MW broadcast signals in a natural subsurface environment and to clarify the differences between the propagation characteristics in this environment and the conventionally studied near-ground environment. The results indicated that the received fields could not be explained solely by surface- or direct-wave models. In particular, this study examined the emergence of $H_z$ and its possible relationship with subsurface displacement currents. The presence of $H_z$ did not indicate the rotation of the polarization plane but represented a redistribution of the magnetic field while preserving the vertically polarized electric field ($E_z$). To interpret these observations, this article introduces the conceptual terms "subground wave" and "rainfall model." These are not new electromagnetic wave theories but descriptive frameworks for the observed field distributions. This study does not replace the classical Zenneck–Sommerfeld theory but complements it by presenting experimentally observed field structures under complex subsurface boundary conditions. Our findings suggest that MW electromagnetic fields can form stable structures that can be received underground. These findings may be applicable to subsurface communication, disaster-resilient systems, and battery-free sensing. Thus, this study provides rare field-based evidence of MW propagation in natural underground environments.

## 2. Materials and Methods

### 2.1 Nanatsuo-guchi quarry

This study was conducted at the Nanatsuo-guchi quarry, a historical underground excavation site for Shakudani Ishi on Mt. Asuwayama, Fukui City, Japan (Figure 1A and 1B). Mt. Asuwayama is a small isolated hill with an elevation of 116.2 m based on GSI benchmark data (GSI, 2026). The quarry is a historical underground excavation site where stone extraction has been performed for approximately 1500 years. Historically, the quarry was also associated with flood control works during the reign of Emperor Keitai. The site consists of dacitic pumiceous volcanic lapilli tuff that formed during the Miocene epoch (Yoshizawa, 2016). This rock exhibits homogeneous properties and contains numerous fine pores that may influence the effective permittivity and conductivity governing electromagnetic wave propagation. The underground excavation space extends several hundred meters into the bedrock. At the entrance, a duralumin tunnel, approximately 2 m wide and 2 m high, was installed as the only metallic structure (Figure 2A and 2B). This section extends approximately 25 m before transitioning into a natural bedrock cavity (Figure 2C and 2D).

Broadcast signal measurements were conducted in a natural bedrock region located well beyond the entrance section. Because the metallic structure was extremely short compared to the wavelength of the MW signals (300–1000 m), the conditions for waveguide resonance or standing-wave formation were not satisfied. Therefore, the influence of the metallic structure on the measurement results was considered local and limited. Thus, the interior was considered a natural bedrock environment with minimal artificial electromagnetic disturbance. The environment was illuminated using incandescent lamps supplied with electricity from the entrance. During the measurements, the power supply was completely disconnected. Because incandescent lamps do not generate electromagnetic noise when unpowered and contain no switching elements, they do not act as noise sources. Preliminary measurements using a spectrum analyzer confirmed the absence of artificial electromagnetic noise and negligible artificial electromagnetic interference, indicating that the environment was electromagnetically quiet. The internal temperature remains approximately 13°C throughout the year, and the humidity is close to 100% (Yoshizawa, 2008). These conditions may promote the formation of thin water films on the bedrock surface, thereby increasing the effective conductivity at the boundary. Therefore, the quarry interior represents a humid, conductive environment rather than a dry cavity. The measurement locations used in this study are hereafter referred to as Points A, B, and C. Point A was defined as an open space approximately 60 m from the entrance (Figure 1B and 2D). Based on topographic data, the bedrock thickness above this point was approximately 20 m. Point B was defined approximately 100 m further inside the quarry, with an estimated overburden of approximately 40 m (Figure 1B and 2E). This increase in bedrock thickness was expected to influence the electric field penetration and magnetic field redistribution. To visualize the geometric relationships between the measurement points, ground surface, and transmitter locations, a three-dimensional terrain model was created using STL data obtained from the Geospatial Information Authority of Japan (GSI). This model was used to visualize the deviations in the arrival directions and the increase in the relative contribution of the $H_z$ component. An additional measurement location, Point C, was established in an open grassy area at the Bunkyo Campus of the University of Fukui for comparison under free-space reception conditions. This site was selected to represent free-space reception conditions without large metal structures or power lines.

### 2.2 Amplitude-modulated (AM) radio broadcasting stations

MW broadcast signals from multiple stations were used to investigate the wave propagation characteristics under different conditions. The NHK Fukui Radio 1 Geba transmitting station (JOFG; 927 kHz; 5 kW) located in Geba, Fukui City, Fukui, Japan, approximately 3.7 km east–southeast of the quarry (Figure 1A), was selected as a nearby source dominated by surface-wave propagation. Its transmitter is a top-loaded vertical steel tower approximately 110 m in height that emits vertically polarized waves. To examine long-distance propagation via ionospheric reflection, signals from the NHK Tokyo Radio 2 Shobu-Kuki transmitting station (JOAB; 693 kHz; 500 kW), located in Shobu, Kuki City, Saitama, Japan, approximately 308 km east of the measurement site, were also used. These signals are typically received through ionospheric reflection, particularly under nighttime conditions. Using nearby and distant transmitters with known signal frequencies, transmission powers, and polarization characteristics, the propagation behavior inside and outside the quarry and its possible underlying mechanisms were analyzed.

### 2.3 Equipment

#### 2.3.1 Radio receiver

A transistor radio receiver (ICF-2001D; Sony, Tokyo, Japan) was used. This receiver offers high sensitivity in the MW band and allows fine gain adjustment. The objective of the measurements was to evaluate the relative variations in the signal strength during antenna rotation rather than the absolute field strength. The receiver was modified to output the signal strength as a direct current (DC) voltage, which was recorded using a data logger (GL980, Graphtec Corporation, Kanagawa, Japan). This configuration enabled continuous data recording synchronized with antenna rotation. The effects of receiver nonlinearity and AGC were confirmed to be negligible for these measurements.

#### 2.3.2 LC resonant loop antenna

An LC resonant loop antenna was constructed using a multiturn coil composed of a 100-strand Litz wire and variable capacitor connected in parallel. This antenna configuration is based on previously developed loop-type radio systems and radio-wave energy harvesting technologies, including the hoop-shaped radio (HOOPRA) and related loop antenna designs (Shoji, 2016; 2023a; 2023b; 2024; 2025a; 2025b). The antenna frame had a diameter of 80 cm, and the resonant frequency was tuned to the target broadcast frequency. The loop antenna was positioned near the internal bar antenna of the receiver to achieve magnetic coupling without a direct electrical connection (Figure 3). A fixed inter-antenna distance of approximately 150 mm was maintained using a resin fixture. The loop plane was oriented such that its normal aligned with the axis of the internal bar antenna. This configuration enhanced the reception sensitivity while enabling stable rotation about the z- and x-axes (Figure 4A).

### 2.3.3 Broadband loop antenna

A broadband loop antenna with a diameter of 80 cm was also constructed. An aluminum pipe formed a loop structure with a gap to maintain the magnetic-loop characteristics. A three-turn Litz wire coil was installed inside the pipe. A balun was connected to the output, and the signal was transmitted via a coaxial cable (5D-2V) to the receiver. This antenna was used to measure the magnetic field components, as loop antennas are primarily sensitive to magnetic fields.

### 2.3.4 Rotation platform and axis definition

A two-axis rotation platform was developed to measure the directional dependence of the received signal strength (Figure 4A). The z-axis was the rotation axis of the loop plane kept vertical and represented the azimuthal arrival direction of the signal. The rotation angle was measured to a 12-bit resolution using a magnetic encoder (AS5600). The orientation at which the loop faced true north, detected using a magnetic compass, was defined as 0°. The signal strength and rotation angle were recorded simultaneously. After determining the direction of the maximum signal strength, the system was reoriented such that the direction of the maximum reception aligned with the z-axis. The x-axis was then defined as the rotation about a horizontal axis, enabling inclination measurements. Figure 4B shows a loop antenna with a loop-plane angle of 0°, whereas Figure 4C shows a loop antenna with a loop-plane angle of 90°. The x-axis angle was measured using the same sensor system. Figure 4D shows the LC resonant loop antenna mounted on a rotation platform.

## 3. Results and Discussion

### 3.1 Geographical characteristics of the Nanatsuo-guchi quarry used for electromagnetic field measurements

The quarry is located beneath Mt. Asuwayama in Fukui City and forms a large subsurface space that extends beneath the central part of the city. The JOFG station (927 kHz), described in Section 2.2, represents a strong nearby source of MW radiation. In addition, three FM broadcasting stations, FBC FM (JOPR, 94.6 MHz), NHK Fukui FM (JOFG-FM, 83.4 MHz), and FM Fukui (JOLU, 76.1 MHz), each with a transmission power of 1 kW, were located on the ridge of Mt. Asuwayama at a distance of approximately 0.6-1.1 km from the quarry. These conditions provide a unique geographical environment for comparing the propagation behavior of radio waves with different wavelengths in and around the quarry.

Assuming free-space propagation, FM broadcast waves are expected to be strongly received near the quarry entrance. However, measurements performed using the ICF-2001D radio receiver showed that the FM signals were strongly attenuated once they entered the bedrock-enclosed entrance section. In practice, audio demodulation cannot be achieved at locations several meters from the entrance. Such signals were considered unreceivable in this study. This blocking behavior was attributed to the combined effects of the geometric boundary conditions at the entrance and attenuation in the surrounding bedrock medium. These observations indicate that VHF-band signals face difficulty when propagating into subsurface spaces. In contrast, the MW broadcast signal of the JOFG station (927 kHz; wavelength of approximately 324 m) was confirmed to be receivable inside the quarry at distances of up to approximately 200 m from the entrance under identical environmental conditions, although the signal attenuation with increasing distance and reception characteristics depended on the antenna conditions. This observation indicates that MW electromagnetic fields can penetrate the surrounding bedrock and reach the interior of the quarry, clearly indicating a significant difference in propagation behavior compared with FM broadcast signals. The observed wavelength dependence of the wave propagation suggests that the attenuation characteristics, transmission under boundary conditions, and resulting spatial field distributions of signals with different frequencies differ significantly.

Considering the short distance between the quarry and the JOFG station, the associated signal was expected to propagate primarily as a surface wave. However, the fact that the signal was received deep inside the bedrock-enclosed quarry indicates that the observed electromagnetic field cannot be fully explained solely by conventional surface-wave propagation models under open-space conditions near the ground surface. Rather, the effective electromagnetic field distribution at the measurement points was modified by the boundary conditions associated with the enclosed subsurface environment. It should also be noted that an electrical power line for lighting was installed inside the quarry. A preliminary investigation confirmed that this conductor could receive MW broadcast signals and reradiate them into the quarry through electromagnetic coupling. When the breaker at the entrance was turned off, this reradiation disappeared, as confirmed by spectral measurements, loss of antenna directivity, and orientation-dependent disappearance of the signal strength. These observations indicate the existence of a reradiation path associated with the incoming power line. To eliminate such artificial influences, all experiments conducted inside the quarry in this study were performed with the breaker turned off.

Under these controlled conditions, the quarry provided a distinctive experimental environment in which the VHF-band FM signals attenuated significantly, but the MW (AM) broadcast signals remained receivable. This allowed for a direct comparison of the wavelength-dependent propagation behavior of the two signal types. The results indicate that MW propagation in subsurface spaces cannot be explained solely by simple leakage or direct intrusion of radio waves. Instead, they suggest that an effective electromagnetic field distribution is formed under the boundary conditions of a closed underground space. Hence, the concept of a subground wave was introduced as a descriptive framework for representing wave propagation behavior in subsurface environments. The subsequent sections present the experimental observations and analyses related to this concept.

**3.2 Radio-wave arrival direction inside and outside the quarry and implications of bedrock medium anisotropy**

To evaluate the reception characteristics of MW broadcast signals inside the quarry, a preliminary survey was conducted from the entrance to the interior. A clear reception of the JOFG signal (927 kHz) was confirmed near the entrance, whereas the reception sensitivity decreased with increasing distance and decreased in deeper sections. The total length of the underground excavation space was approximately 340 m. For quantitative analysis, Point A was defined as a location approximately 60 m from the entrance (Figure 1B and 2D), and Point B was defined as a location approximately 150 m from the entrance (Figure 1B and 2E). Reference measurements were first conducted at measurement point C on the Bunkyo Campus of the University of Fukui outside the quarry. A broadband loop antenna was rotated around the z-axis to determine the arrival direction and around the x-axis to evaluate the angular dependence of the reception sensitivity. The signal strength of JOFG (927 kHz) and the rotation angle were recorded simultaneously at point C (Figure 5A). The maximum reception sensitivity was observed at approximately 155° (true north = 0°), which is consistent with the direction of the transmitting station. This confirms that the measurement system reliably captures the arrival direction under free-space conditions. Fixing the *z-axis* along the direction of maximum reception, the antenna was rotated around the *x-axis* (Figure 5B). Under free-space conditions, the magnetic field was dominated by $H_x$. Reception sensitivity is maximized when the loop plane is oriented to receive this component. The observed angular dependences of the reception sensitivities of the loop and built-in bar antennas were consistent, confirming that the measurement system detected the magnetic field components in the MW signals.

The measurements were then conducted at point A (Figure 2D) inside the quarry. When the loop plane was kept vertical and rotated around the z-axis, the reception sensitivity was maximized at an azimuth angle of approximately 90° (Figure 6A). Because the JOFG station was located in the east–southeast direction (approximately 120°), the observed arrival direction shifted northward relative to the transmitter direction. This deviation indicates that propagation in the subsurface environment is not governed solely by free-space geometry. Instead, radio waves reached the measurement point through the surrounding bedrock, and the effective transmission path was influenced by spatial variations in bedrock thickness. In addition, the reception sensitivity at point A remained high over a wide angular range rather than forming a sharp peak. This behavior indicates that the electromagnetic field at the measurement point was not dominated by a single propagation path, but reflected a distributed field structure formed under subsurface boundary conditions. Similar measurements were conducted at point B (Figure 2E), which was located deeper inside the quarry. The bedrock thickness above this point was approximately 40 m, which was greater than that at point A. The maximum reception sensitivity was observed at an azimuth angle of approximately 130° (Figure 7A), corresponding to a direction closer to the transmitter than that observed at point A. Furthermore, the region of high reception sensitivity extended over a wide angular range, indicating reduced directivity.

These results indicate that, as the bedrock thickness increased, propagation paths with weak attenuation became dominant, leading to an apparent shift in the arrival direction toward the transmitter. Furthermore, the distribution of the maximum reception sensitivity over a wide angular range suggests an increased contribution of distributed propagation paths through the bedrock. Such behavior cannot be explained by simple free-space propagation. Instead, it reflects the influence of the geometric structure of the quarry and the directional dependence of the electromagnetic wave properties in the bedrock. In this study, bedrock medium anisotropy refers to spatial variations in the effective conductivity, permittivity, and loss characteristics of the bedrock owing to moisture distribution and structural heterogeneity. Therefore, the observed deviations in the arrival direction and changes in the angular distribution of the reception sensitivity can be attributed to the anisotropy pertaining to bedrock thickness and subsurface boundary conditions. These findings indicate that MW propagation in subsurface environments is governed by a complex interaction between the geometry and properties of the bedrock medium, resulting in an effective electromagnetic field distribution that is distinct from that observed in conventional surface-wave propagation.

### 3.3 Discovery of the $H_z$ component from changes in the plane angle and reception characteristics of the loop antenna

The azimuth corresponding to the maximum reception sensitivity determined in the previous section was fixed, and the plane of the broadband loop antenna was rotated about the *x-axis* to evaluate the angular dependence of the magnetic field reception sensitivity. The measurements were conducted at Point A receiving a JOFG signal (927 kHz) while tilting the plane from 0° (horizontal) to 90° (vertical). The relative signal strengths were recorded (Figure 6B). The resulting angular dependence of the reception sensitivity did not exhibit a simple sinusoidal pattern. Instead, an asymmetric distribution was observed, which was characterized by a clear dip in reception sensitivity near 40°–50° and a broad maximum around 120°–130°, instead of a sharp maximum. In addition, the signal was received even when the loop plane was maintained in a horizontal orientation. Furthermore, the signal strength showed a nonmonotonic variation, first decreasing and then increasing with increasing tilt angle. These characteristics differed significantly from the symmetric dipole-like response expected under conventional surface-wave propagation. These observations indicate that the magnetic field at a measurement point cannot be described by a single dominant component. Instead, the results indicated that *the* $H_x$ and $H_z$ components coexisted, forming a composite field. A dip in the reception sensitivity near 45° could be interpreted as a reduction in the effective magnetic flux when the contributions of $H_x$ and $H_z$ became comparable in the direction normal to the loop plane.

A similar measurement was conducted at Point B, which was located deeper inside the quarry (Figure 7B), where the bedrock thickness was greater than that at Point A. The angular dependence of the reception sensitivity exhibited a more pronounced asymmetry. The reception sensitivity decreased significantly near 100°–110°, whereas the maximum reception sensitivity was observed at loop-plane angles of 30° or less, corresponding to a nearly horizontal loop orientation. This shift of the peak toward smaller tilt angles indicated that the relative contribution of $H_z$ increased in the deeper regions of the quarry. This interpretation was further supported by additional reception tests. When the loop antenna was placed directly on the ground, a high reception sensitivity was achieved for the horizontal orientation. Moreover, measurements using the built-in bar antenna of the receiver showed that the maximum reception sensitivity inside the quarry was observed when the long axis of the bar antenna was oriented vertically. This behavior contrasted with that observed outside the quarry, where the maximum reception sensitivity was observed in the horizontal orientation. These results consistently indicate that the magnetic field distribution inside the quarry differs from that near the ground surface, and the $H_z$ component becomes significant in subsurface environments.

The reception behavior of the JOAB (693 kHz) signals arriving via ionospheric reflection inside the quarry was also examined. Because these signals are weak and susceptible to fading, enhancement of reception sensitivity was necessary. Under the present measurement configuration, the built-in bar antenna exhibited a higher reception sensitivity than the broadband loop antenna connected as an external antenna. To further enhance reception sensitivity, the built-in bar antenna was magnetically coupled to an LC resonant-loop antenna. To confirm the measurement setup, reference measurements were first conducted outside the quarry at Point C using a JOFG signal (927 kHz). The observed arrival direction of the JOFG was in good agreement with the actual transmitter direction of approximately 155° (Figure 8A). The results exhibited a symmetric angular dependence of reception sensitivity, with the maximum value observed near 90° and the minimum value observed near 0° (Figure 8B), consistent with the dominance of $H_x$ in the classical surface-wave model. Based on this confirmation, reception measurements of JOAB were performed. The observed arrival direction of JOAB was in good agreement with the due-east direction (i.e., 90°) (Figure 8C). The results exhibited a symmetric angular dependence of reception sensitivity, with the maximum value observed near 90° and the minimum value observed near 0° (Figure 8D), consistent with the dominance of $H_x$ under ionospheric-reflection conditions.

A similar measurement was conducted at Point A inside the quarry. The ionospherically reflected JOAB signal (693 kHz) exhibited a distinct asymmetric pattern, with reception sensitivity dips near 40°–50° and a high reception sensitivity distributed over a wide angular range (Figure 9A). The surface-wave JOFG signal (927 kHz) likewise exhibited a distinct asymmetric pattern, with a dip in reception sensitivity appearing near 40°–50°, coinciding with that observed for the ionospherically reflected JOAB signal, and with a high reception sensitivity distributed over a wide angular range (Figure 9B). The observed patterns were reproducible across multiple measurement locations and could not be attributed to measurement artifacts. These observations provided experimental evidence that $H_x$ is dominant near the ground surface, whereas the relative contribution of $H_z$ becomes significant in subsurface environments. These findings indicate that the MW electromagnetic fields in the quarry formed a composite magnetic field structure distinct from that expected in conventional surface-wave propagation and provided a physical basis for the conceptual framework introduced in this study.

### 3.4 Verification of the reception characteristics of MW broadcast waves inside the quarry based on magnetic field coupling using an LC resonant loop antenna

Shakudani Ishi is a dacitic pumiceous volcanic lapilli tuff with a porous structure and high water content. Its magnetic permeability is nearly equal to that of free space, and as a rock medium, it strongly attenuates VHF-band electromagnetic waves. However, as demonstrated in this study, MW broadcast signals penetrated the bedrock beneath Mt. Asuwayama and remained receivable inside the quarry. This observation indicates that MW electromagnetic fields can propagate through subsurface media while undergoing attenuation. The observational results revealed an electromagnetic field behavior that could not be fully explained by conventional MW propagation models based on surface waves, such as Zenneck, Sommerfeld, and Norton waves. In particular, the presence of $H_z$ inside the quarry differed from the field structure generally assumed near the ground surface, where $H_x$ is dominant. The observed $H_z$ component was interpreted as a redistribution of the magnetic field under subsurface boundary conditions, whereas the electric field remained vertically polarized ($E_z$). Therefore, the emergence of $H_z$ does not indicate a rotation of the polarization plane but reflects a change in the relative contributions of the magnetic field components.

Based on these findings, this study introduced the concept of a subground wave as a descriptive framework. In this context, a subground wave refers to an MW electromagnetic field that penetrates bedrock and forms a subsurface field structure in which $H_z$ becomes significant. This concept is consistent with the conventional surface-wave theory and extends the interpretation of the theory to subsurface environments, where boundary conditions differ from those of open space. The observed $H_z$ component could be considered to originate from field redistribution under the influence of varying geometry and properties of the bedrock. Examples include bedrock property anisotropy, moisture distribution variations, and electromagnetic field confinement within quarry structures. The experimental results, particularly the nonmonotonic angular dependence of reception sensitivity observed in loop-antenna measurements, indicate that the magnetic field cannot be described by a single component but should be interpreted as a composite structure composed of the $H_x$ and $H_z$. The observed minimum reception sensitivity occurred when these components contributed comparably in the direction normal to the loop plane, resulting in a reduction in the effective magnetic flux. To interpret this behavior, a rainfall model was proposed as a conceptual framework. In this model, a vertically polarized electric field ($E_z$) induces currents ($J$) near the ground surface under subsurface boundary conditions. These currents generate the vertical magnetic field component $H_z$, accompanied by the formation of the horizontal electric field component $E_x$. This coupling process represents the redistribution of the electromagnetic field inside the quarry. In this framework, the components $H_{x,q}$ and $E_{z,q}$ represent the redistributed field components formed inside the quarry. The relation $S_z = E_z \times H_x$ represents the vertical component of the Poynting vector associated with the observed field redistribution and provides a qualitative interpretation of the electromagnetic energy flow. These relationships are schematically illustrated in Figure 10.

The observational results, including the attenuation of $H_x$, the increasing magnitude of $H_z$ with increasing depth, and the characteristic angular dependence of the reception sensitivity, are qualitatively consistent with this model. These findings indicate that electromagnetic field distributions in subsurface environments are governed by interactions among boundary conditions, medium properties, and field-coupling processes. A similar behavior has been reported in tunnel environments, where electric fields are strongly attenuated while magnetic fields penetrate more effectively. In such cases, the $H_z$ component can be detected, supporting the interpretation that subsurface propagation involves field redistribution rather than simple transmission. Despite the different physical structures of the tunnels and quarry, these observations are consistent with the subground wave concept proposed in this study.

Overall, the results demonstrate that MW electromagnetic fields can propagate in subsurface environments while forming characteristic field distributions distinct from those observed near the ground surface. The subground wave concept and rainfall model represent descriptive frameworks for interpreting these observations. Further work is required to develop quantitative models and numerical simulations incorporating the geological structure, medium properties, and propagation conditions.

## 4. Conclusion

In this study, the propagation characteristics of MW (AM) radio broadcast signals were empirically investigated in a subsurface space, specifically at the Nanatsuo-guchi quarry of the Shakudani Ishi excavation site located on Mt. Asuwayama, Fukui City, Japan. This environment provides a unique combination of bedrock-enclosed underground spaces and proximity to MW broadcasting stations. The results demonstrated a strong wavelength dependence of subsurface wave propagation. VHF-band FM broadcast waves were rapidly attenuated within several tens of meters from the quarry entrance, whereas MW signals penetrated the bedrock and remained receivable even at deeper locations. This contrast highlights the distinctive ability of MW electromagnetic fields to propagate in subsurface environments. Arrival-direction measurements and tilt-angle dependence of loop antennas revealed that, in addition to the $H_x$ component assumed in conventional surface-wave propagation, the $H_z$ component was prominently observed inside the quarry. The observed angular characteristics, including a reception sensitivity minimum near 40°–50°, indicated that the magnetic field could not be described by a single component but instead reflected a composite magnetic field structure consisting of the $H_x$ and $H_z$, whose relative contributions varied depending on the subsurface boundary conditions.

These findings cannot be fully explained by conventional surface-wave models, such as Zenneck, Sommerfeld, and Norton waves; however, they remain consistent with the fundamental framework of electromagnetic wave theory. The observed behavior was interpreted as a redistribution of the magnetic field while preserving the transverse nature of electromagnetic waves. In this framework, the redistributed components $H_{x,q}$ and $E_{z,q}$ were interpreted as the characteristic field components formed inside the quarry under subsurface boundary conditions. Based on these observations, this study introduced the concept of a subground wave as a descriptive framework for understanding MW propagation in subsurface environments. In addition, a rainfall model was proposed to describe the coupling process in which a vertically polarized electric field ($E_z$) induces currents ($J$) under subsurface boundary conditions, generating the $H_z$ component together with the formation of horizontal electric fields ($E_x$).

These results provide a new perspective that complements the conventional surface-wave theory and contributes to the understanding of electromagnetic propagation in complex environments. Future work will focus on quantitative modeling and numerical simulations that incorporate geological structures and medium properties. This study represents an initial step toward understanding MW propagation in subsurface environments, with potential applications in underground communication, disaster-resilient information transmission, and subsurface electromagnetic wave observation.

**Acknowledgments**

The author expresses their sincere gratitude to Mr. Yoshiei Fukushima, President of Echizen Ishi Co., Ltd., for preserving and managing the Nanatsuo-guchi quarry, permitting its use as an experimental field, and for supporting this research. The author also thanks Ms. Naoko Shoji for assistance with equipment transport, on-site measurements, and valuable discussions. This work was supported by JSPS KAKENHI Grant Number JP24K15388 and an FY2024 research grant from the Research and Education Center for Regional Environment at the University of Fukui.

**Conflict of Interest**

The author declares no conflicts of interest related to this study.

**Data Availability Statement**

Due to the nature of the field measurements and the experimental environment, the data are not publicly available but may be available from the corresponding author upon reasonable request.

**Legend of Figures**

Figure 1

Geographical relationship between the measurement points and broadcasting stations around Mt. Asuwayama and the internal structure of the quarry.

(A) Spatial arrangement of measurement locations and broadcasting stations. White circles indicate the measurement points. The Nanatsuo-guchi quarry entrance is located on Mt. Asuwayama. FM stations (FBC FM, 94.6 MHz, JOPR; NHK Fukui FM, 83.4 MHz, JOFG-FM; and FM Fukui, 76.1 MHz, JOLU) are located near the ridge, and the AM station JOFG (927 kHz) is located approximately 3.7 km away. The dashed line indicates the alignment between the JOFG station and quarry entrance.

(B) Schematic plan view of the quarry showing measurement Points A and B.

Figure 2

Observation environment in the Nanatsuo-guchi quarry. The quarry consisted of a large bedrock cavity where electromagnetic measurements were conducted.

(A) Entrance view of the quarry from the outside.

(B) Entrance view of the quarry from inside.

(C) Interior view near the entrance.

(D) Measurement being conducted at Point A.

(E) Measurement being conducted at Point B.

Figure 3

Schematic of the receiving system consisting of an LC resonant loop antenna (diameter: 800 mm) and a modified radio receiver (ICF-2001D). (A) Side view. (B) Front view. The loop antenna and internal bar antenna were separated by 150 mm. The system was rotated to measure the angular dependence of the reception sensitivity.

Figure 4

Measurement setup and observation configuration.

(A) Two-axis rotation platform enabling azimuth (z-axis) and tilt (x-axis) rotations.

(B) Example of a broadband loop antenna with a loop-plane angle of 0°.

(C) Reception of the JOFG signal (927 kHz) using a broadband loop antenna with a loop-plane angle of 90° outside the quarry.

(D) Reception of the ionospherically reflected JOAB signal (693 kHz) using an LC resonant loop antenna with a loop-plane angle of 90° at Point C on the Bunkyo Campus of the University of Fukui.

Figure 5

Relationship between the received voltage (RV) and antenna angle measured using a broadband loop antenna at the Bunkyo Campus of the University of Fukui while receiving the JOFG signal (927 kHz).

(A) Signal variation with azimuth angle (AA; z-axis rotation).
(B) Signal variation with loop inclination angle (LA; x-axis rotation).

Figure 6
Relationship between the received voltage (RV) and antenna angle measured using the broadband loop antenna at Point A inside the quarry while receiving the JOFG signal (927 kHz).
(A) Signal variation with azimuth angle (AA).
(B) Signal variation with loop inclination angle (LA).
The arrival direction of the signal and the angular dependence of the reception sensitivity inside the quarry are clearly identified.

Figure 7
Relationship between the received voltage (RV) and antenna angle measured using the broadband loop antenna at Point B inside the quarry while receiving the JOFG signal (927 kHz).
(A) Signal variation with azimuth angle (AA).
(B) Signal variation with loop inclination angle (LA).
Distinct changes in the angular dependence of the reception sensitivity were observed in the deeper regions of the quarry.

Figure 8
Angular dependence of the received voltage (RV) measured using the LC resonant loop antenna at Point C on the Bunkyo Campus of the University of Fukui.
(A) Azimuth-angle (AA) dependence of the surface-wave signal from JOFG (927 kHz).
(B) Loop inclination angle (LA) dependence of the surface-wave signal from JOFG (927 kHz).
(C) Azimuth-angle (AA) dependence of the ionospherically reflected signal from JOAB (693 kHz).
(D) Loop inclination angle (LA) dependence of the ionospherically reflected signal from JOAB (693 kHz).

Figure 9
Loop inclination angle (LA) dependence of the received voltage (RV) measured using the LC resonant loop antenna at Point A inside the quarry.
(A) Ionospherically reflected signal from JOAB (693 kHz).
(B) Surface-wave signal from JOFG (927 kHz).

Figure 10
Conceptual schematic of the “rainfall model” describing the redistribution of electromagnetic fields inside the quarry. A vertically polarized electric field ($E_z$) induces horizontal electric fields ($E_x$) and associated currents ($J = \sigma E_x$) near the ground surface, leading to the generation of a vertical magnetic field component ($H_z$) in the subsurface space. The components $H_{x,q}$ and $E_{z,q}$ represent the redistributed field components within the quarry. $S_z = E_z \times H_x$ represents the vertical component of the Poynting vector associated with the incident electromagnetic field. This figure provides a qualitative interpretation of the observed magnetic field distribution.

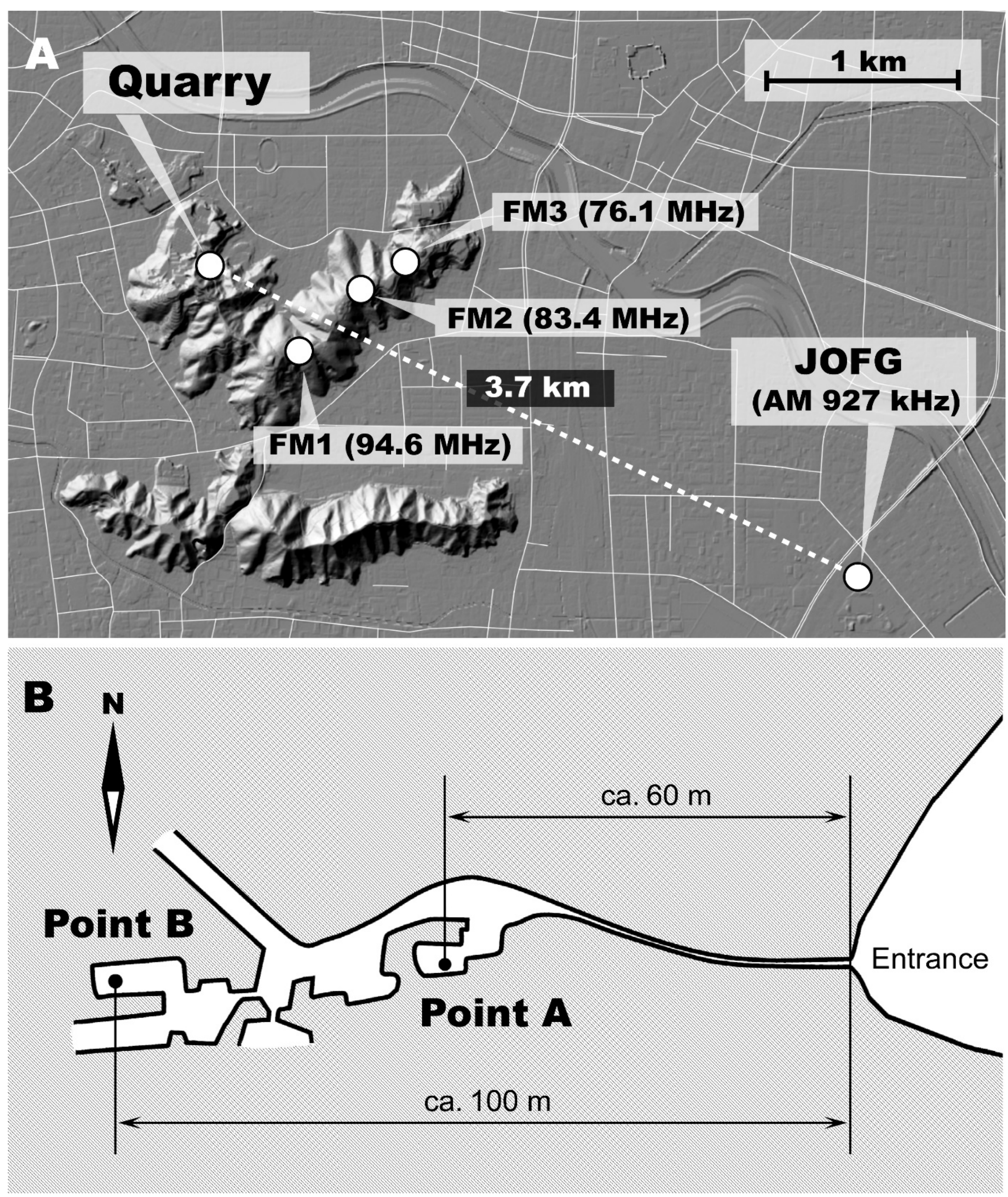

**Figure 1**
**Eiichi Shoji**

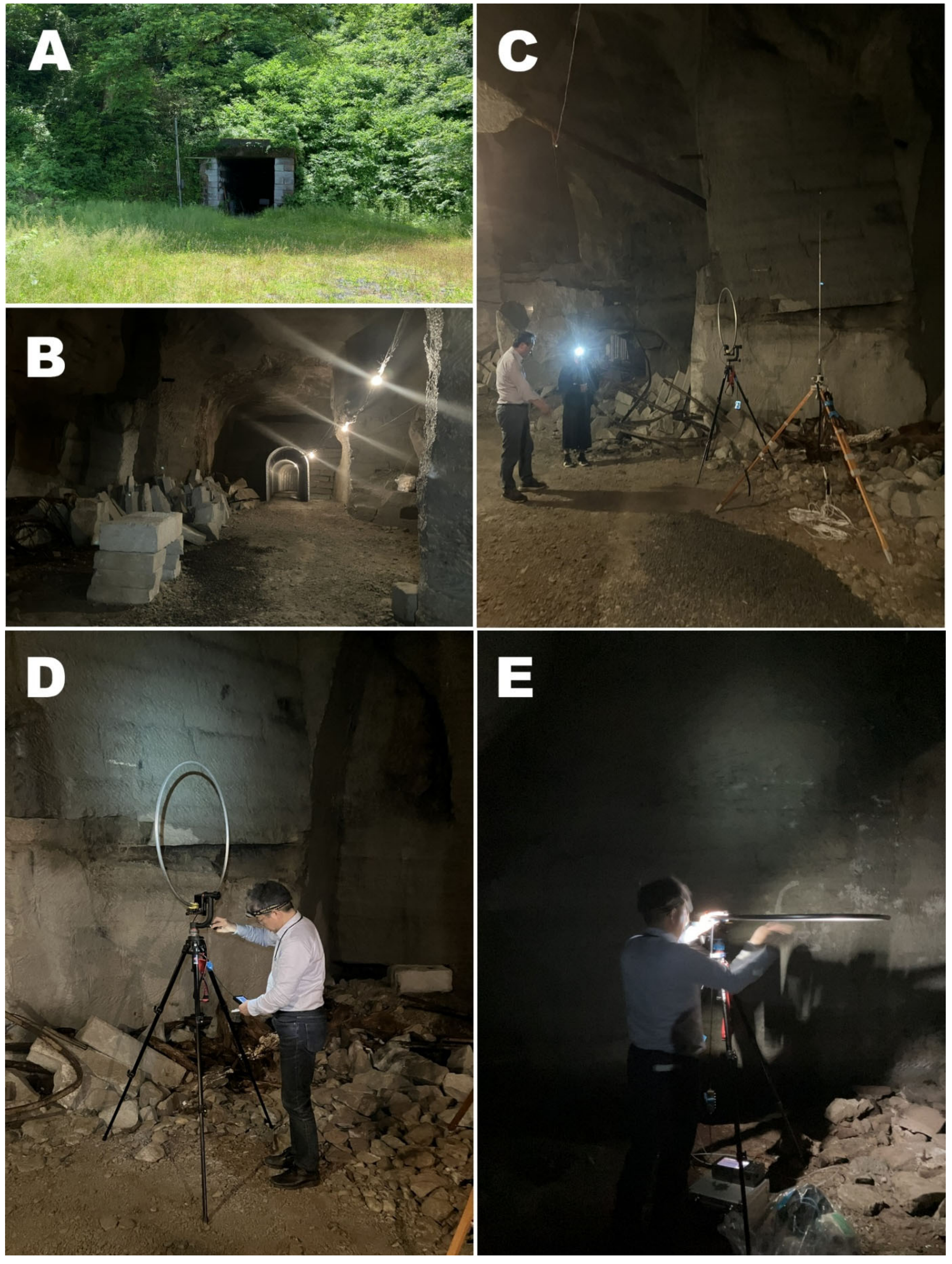

**Figure 2**
**Eiichi Shoji**

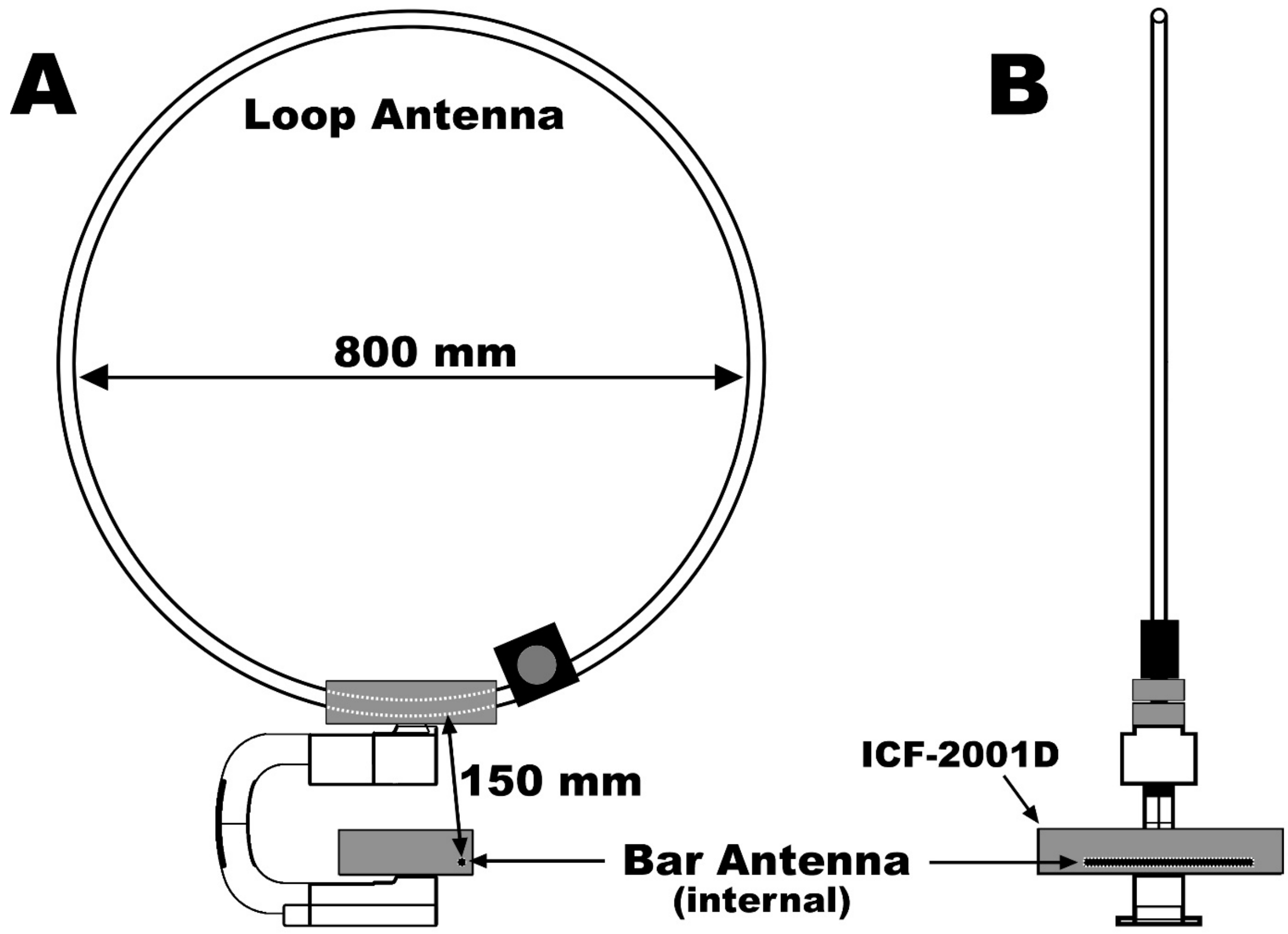

**Figure 3**

**Eiichi Shoji**

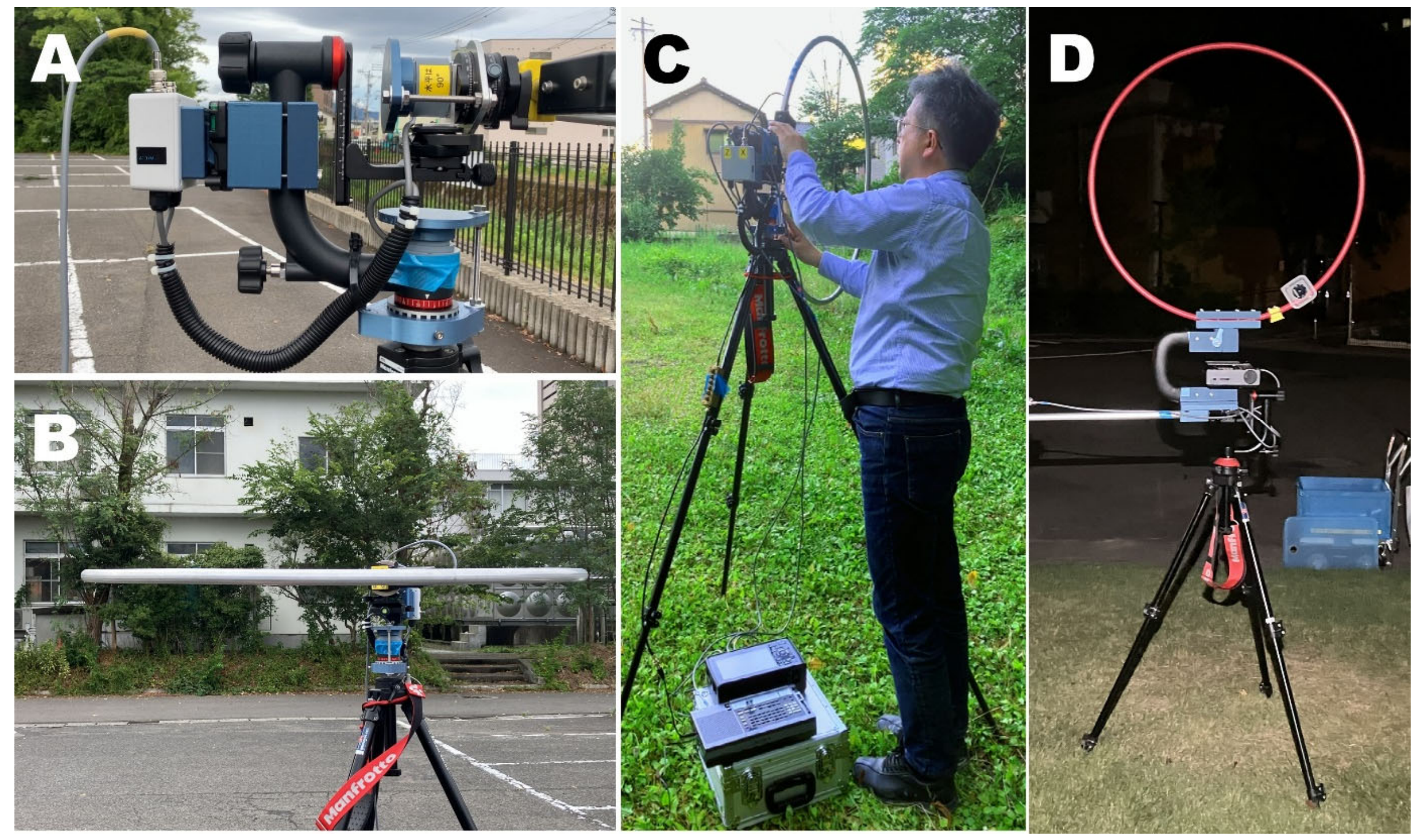

**Figure 4**

**Eiichi Shoji**

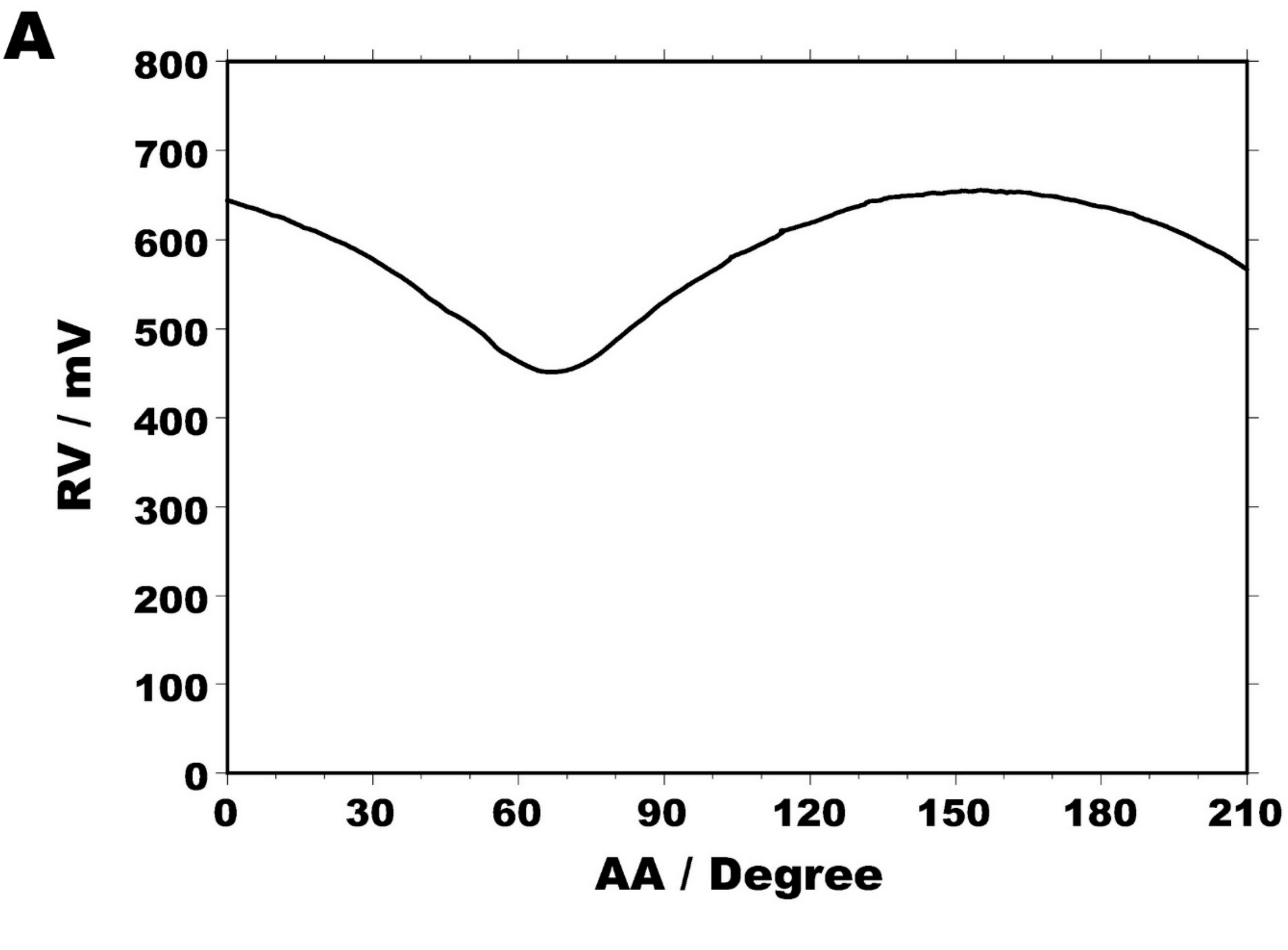

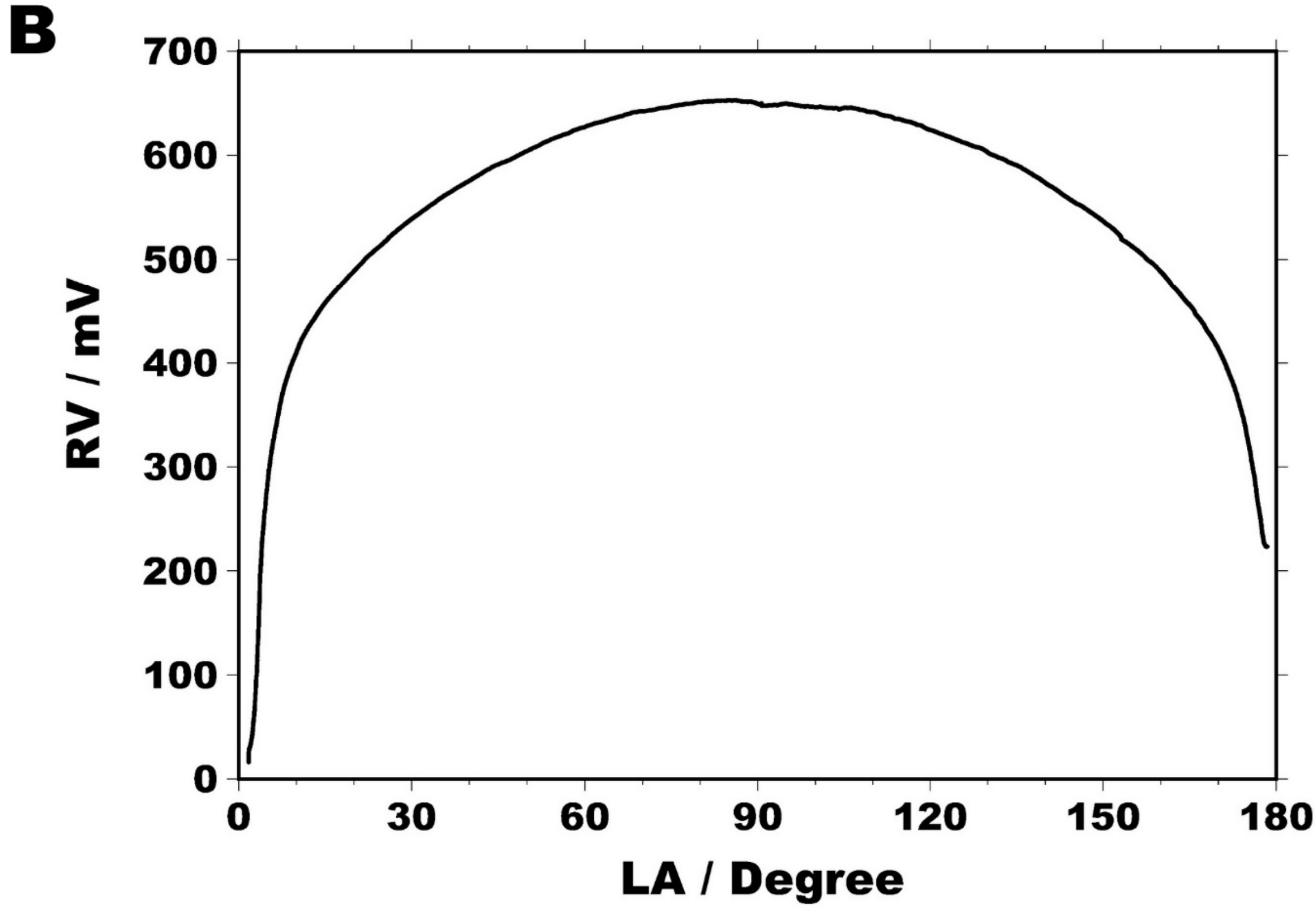

**Figure 5**

**Eiichi Shoji**

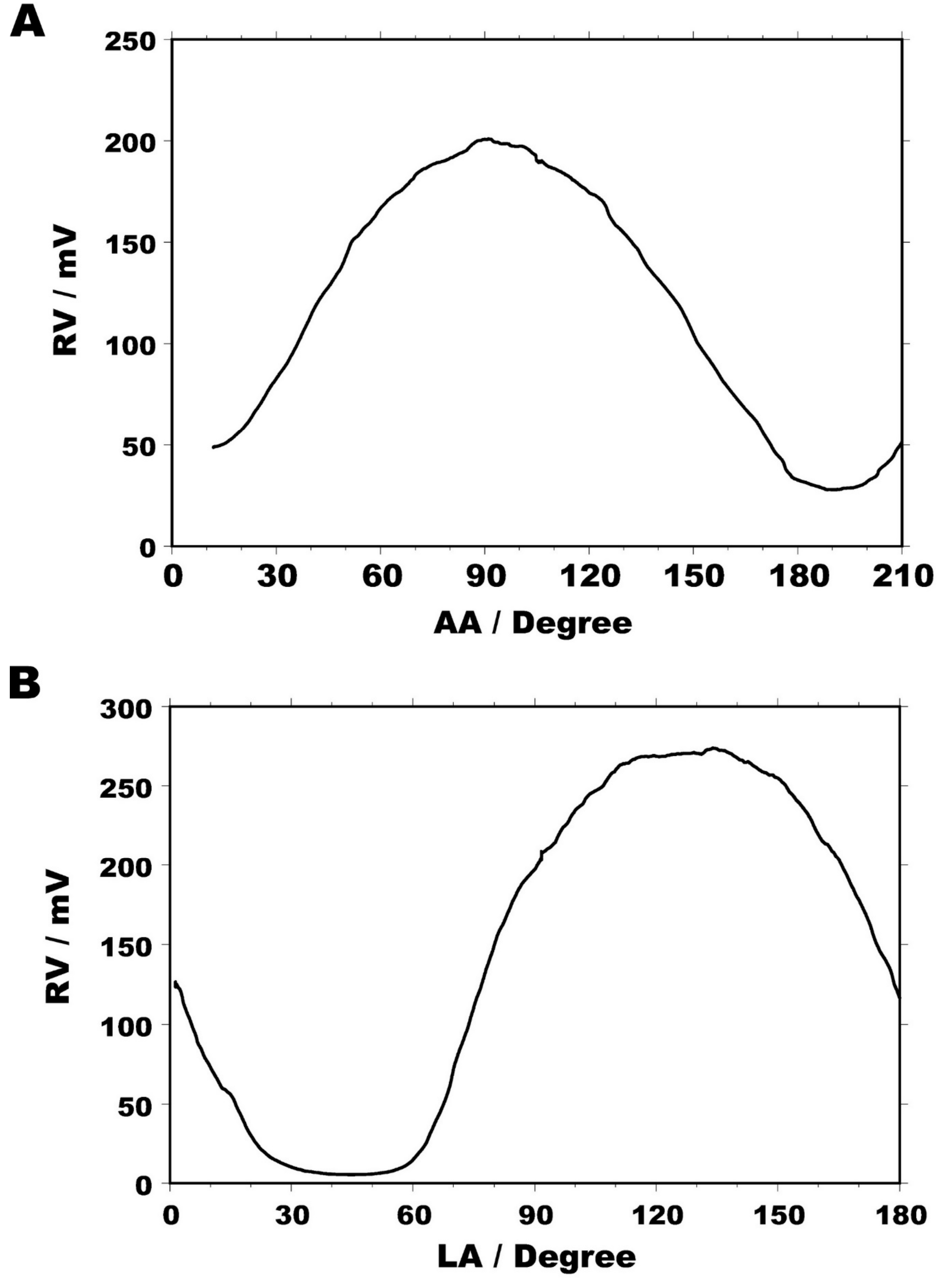

**Figure 6**

**Eiichi Shoji**

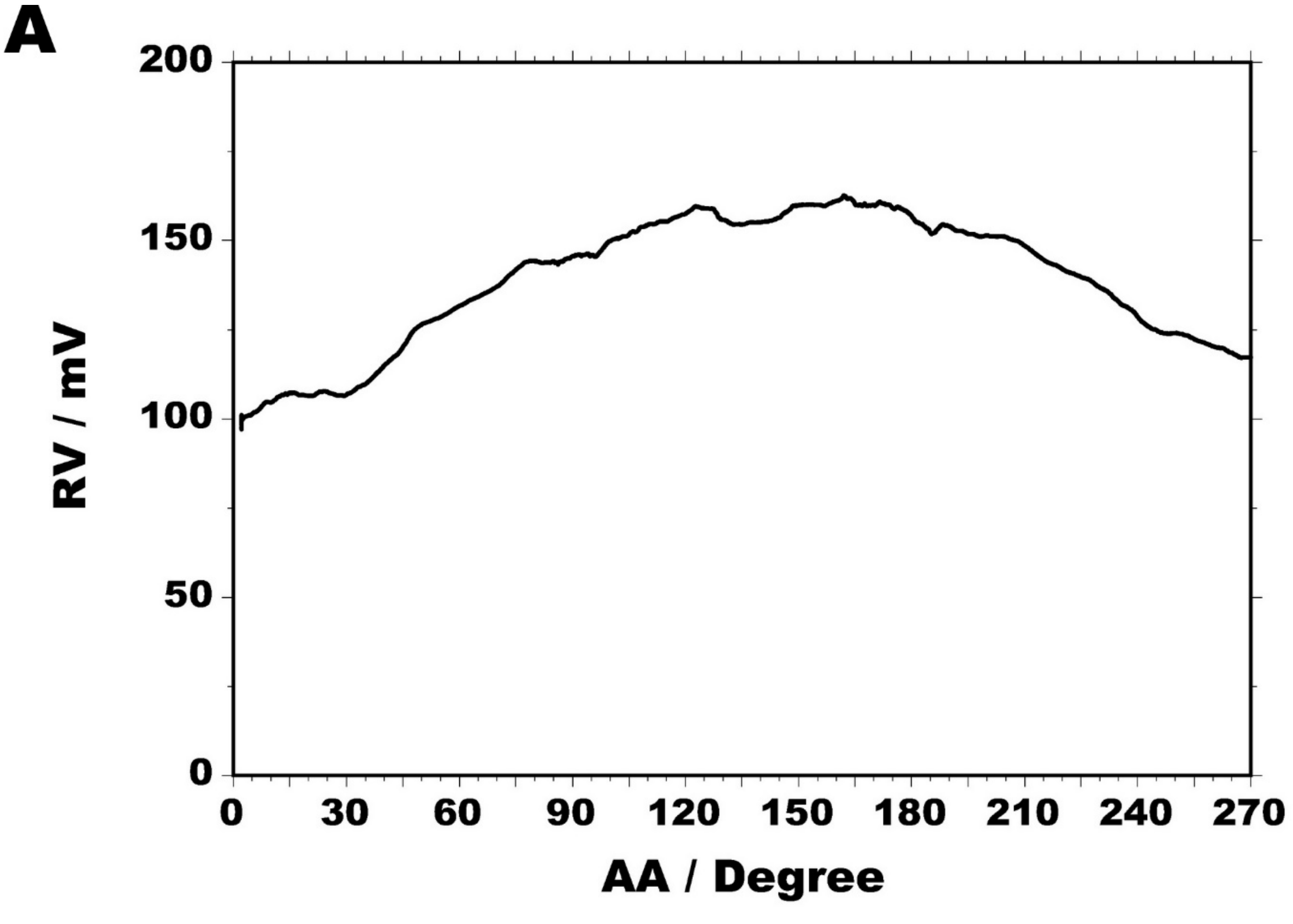

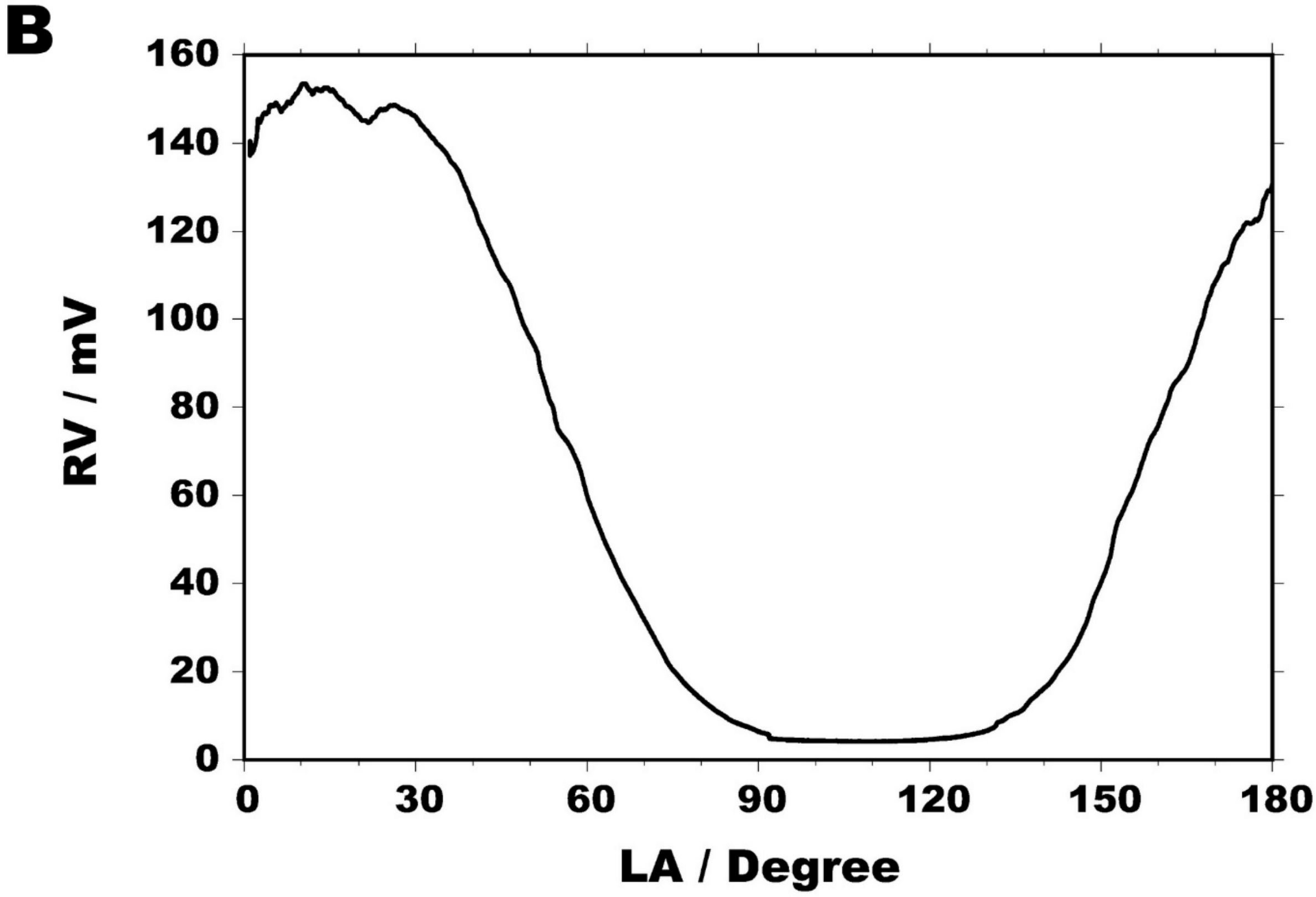

**Figure 7**

**Eiichi Shoji**

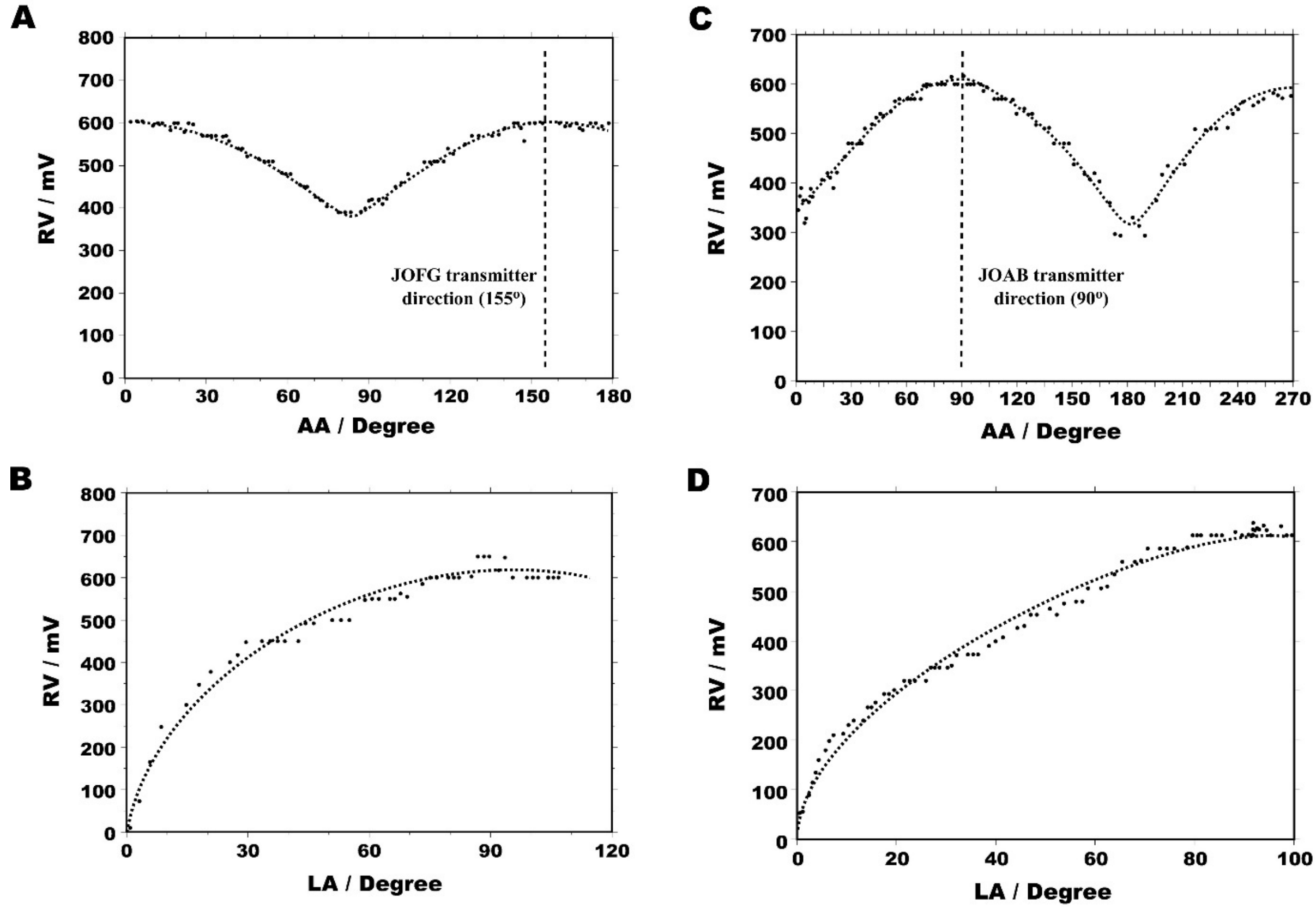

**Figure 8**

**Eiichi Shoji**

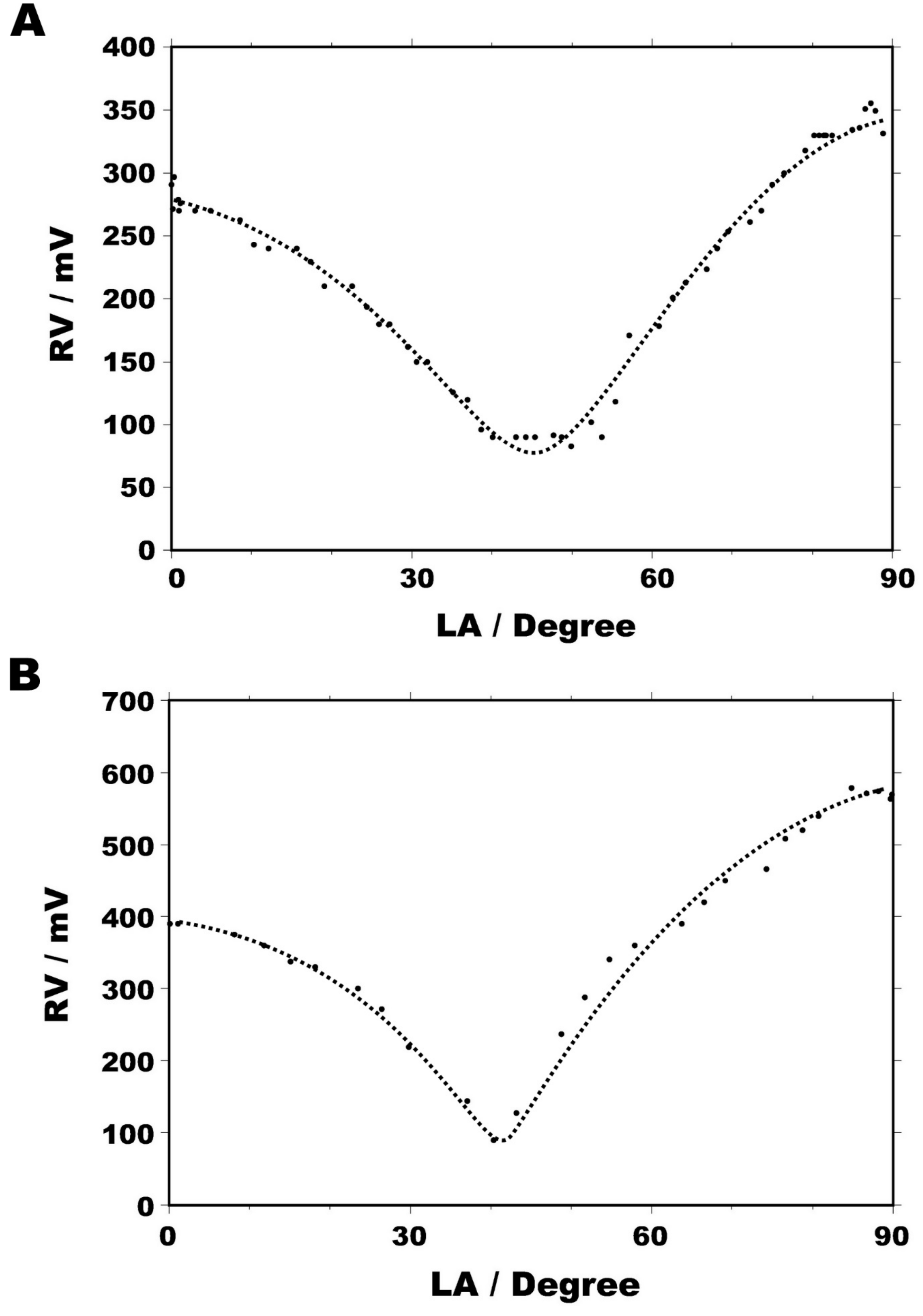

**Figure 9**

**Eiichi Shoji**

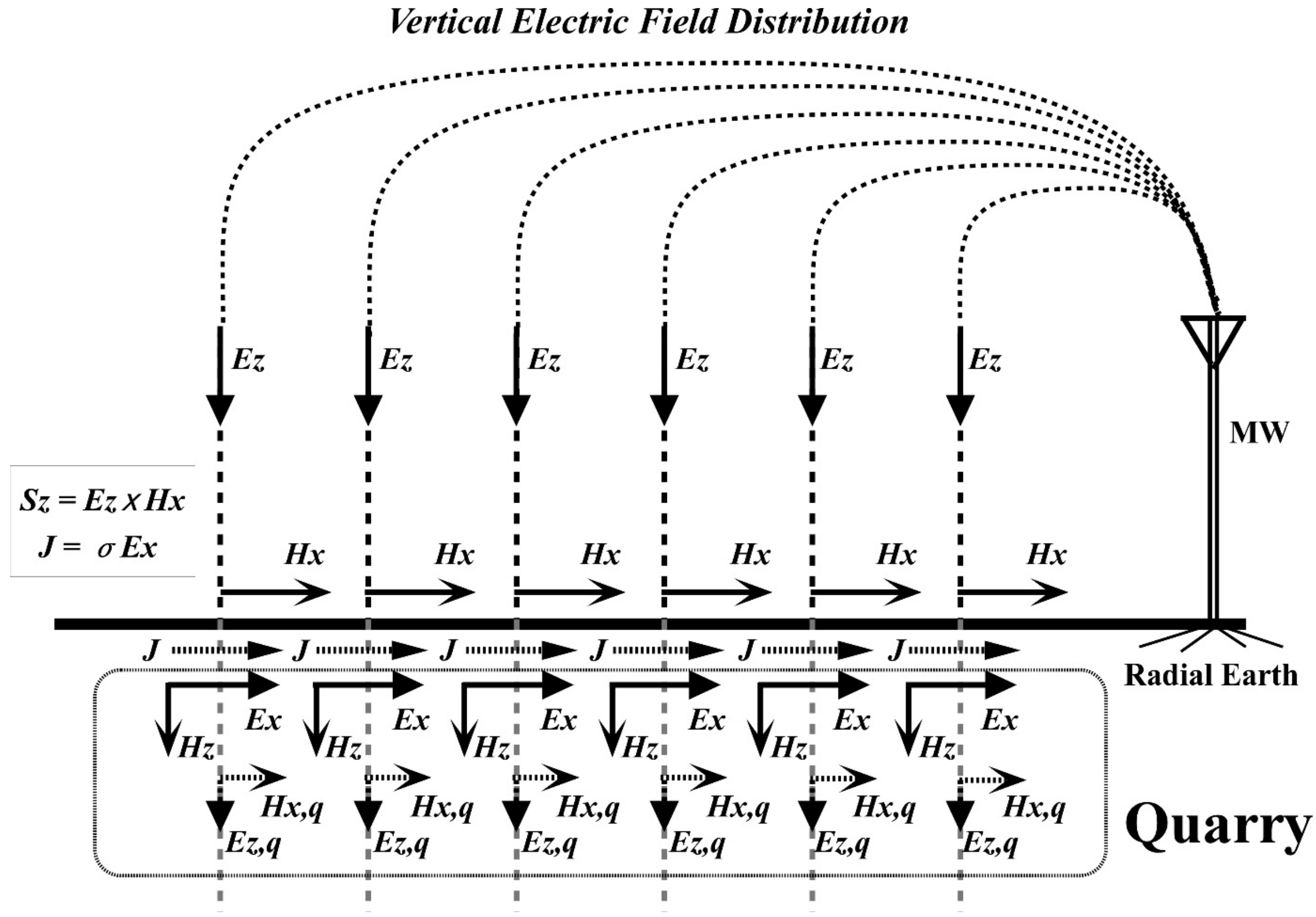

**Figure 10**

**Eiichi Shoji**